\documentclass[a4paper,12pt,abstract=true,numbers=noenddot]{scrartcl}
\pdfoutput=1
\usepackage{graphicx}
\usepackage{pdfpages}
\usepackage{amsmath,verbatim,mathrsfs,array,layout,bbold,textcomp,amssymb,latexsym}
\usepackage[titletoc,title]{appendix}
\usepackage[affil-it ]{authblk}

\usepackage{hyperref}
\usepackage{xcolor}
\usepackage{braket}
\usepackage{subfig}
\usepackage{cancel}
\usepackage{cite}

\hypersetup{
    colorlinks,
    linkcolor={red!50!black},
    citecolor={blue!50!black},
    urlcolor={blue!80!black}
}

\newcommand{\CL}{{\rm C.L.}}

\def\beqa{\begin{align}}
\def\eeqa{\end{align}}
\def\beq{\begin{equation}}
\def\eeq{\end{equation}}

\def\lsim{\mathrel{\rlap{\lower4pt\hbox{\hskip1pt$\sim$}}
     \raise1pt\hbox{$<$}}}
\def\gsim{\mathrel{\rlap{\lower4pt\hbox{\hskip1pt$\sim$}}
     \raise1pt\hbox{$>$}}}

\setcapindent{1em}
\setkomafont{captionlabel}{\bf}
\setkomafont{caption}{\it}

\begin{document}
\renewcommand\Authands{, }
\unitlength = 1mm
\title{Large BR($\boldsymbol{h\to\tau\mu}$) in the MSSM?}
\date{\today}
\author{Daniel Aloni\thanks{\texttt{daniel.aloni@weizmann.ac.il}}}
\author{Yosef Nir\thanks{\texttt{yosef.nir@weizmann.ac.il}}}
\author{Emmanuel Stamou\thanks{\texttt{emmanuel.stamou@weizmann.ac.il}}}

\affil{Department of Particle Physics and Astrophysics,
Weizmann Institute of Science, Rehovot 7610001, Israel}

\maketitle

\begin{abstract}
\noindent
We study how large the rate of the lepton-flavor violating Higgs decay $h\to\tau\mu$ can be in the (R-parity conserving) MSSM.
We make no assumptions, such as universality or alignment, about the flavor structure of the MSSM.
We only assume that all couplings and, in particular, the trilinear scalar ones, are perturbative.
We take into account lower bounds on the bino and slepton masses from $\tau\to\mu\gamma$ and $h\to\gamma\gamma$
as well as upper bounds on the trilinear scalar couplings from the requirement that the global minimum
is not charge breaking.
We find that in highly fine-tuned regions of parameter space, the ratio ${\rm BR}(h\to\tau\mu)/{\rm BR}(h\to\tau\tau)$
can be enhanced by about three orders of magnitude above the estimate from naive dimensional analysis,
but still about two orders of magnitude below the present bound.
Thus, if $h\to\tau\mu$ is experimentally established to be close to present bounds, the MSSM will be excluded.
\end{abstract}

\section{Introduction}
\label{sec:int}
The first direct searches for the lepton-flavor violating (LFV) Higgs
decay $h\to\tau\mu$ were carried out by the CMS and ATLAS collaborations
\cite{Khachatryan:2015kon,Aad:2015gha} yielding the upper bounds:
\begin{equation}\label{eq:BRmutauu}
{\rm BR}(h\to\tau\mu)<\left\{\begin{matrix}1.51\times10^{-2}&{\rm CMS,}\\ 1.85\times10^{-2}&{\rm ATLAS,} \end{matrix}\right.
\end{equation}
and the ranges:
\begin{equation}\label{eq:BRmutauc}
{\rm BR}(h\to\tau\mu)=\left\{\begin{matrix}(8.4^{+3.9}_{-3.7})\times10^{-3}&{\rm CMS,}\\
(7.7\pm6.2)\times10^{-3}&{\rm ATLAS.} \end{matrix}\right.
\end{equation}
The $h\to\tau\mu$ decay has several aspects that are worth emphasizing:
\begin{itemize}
\item It violates the lepton-flavor symmetry $U(1)_\mu\times U(1)_\tau$,
      which is an accidental symmetry of the Standard Model (SM).
\item It is a flavor changing neutral current (FCNC) process.
\item It violates the prediction that the Yukawa matrix is proportional to the mass
      matrix, $Y^E\propto M_E$, which applies at the tree level to all models of
      Natural Flavor Conservation (NFC).
\end{itemize}
Due to these three aspects, an observation of $h\to\tau\mu$ in present experiments
will have far reaching implications.

In this work we ask whether an $h\to\tau\mu$ decay rate close to the
near future sensitivity of the LHC experiments, ${\rm BR}(h\to\tau\mu)={\cal O}(0.01)$,
can be accounted for by the minimal supersymmetric standard model (MSSM).
The branching ratio depends on the total width of the
Higgs, which is experimentally unknown and which, within the MSSM, depends on the entire supersymmetric spectrum.
To avoid the dependence on sectors unrelated to LFV and
on experimentally yet-unconstrained observables, we consider the ratio of
branching ratios,
\begin{equation}
R_{\tau\mu/\tau\tau}\equiv\frac{{\rm BR}(h\to\tau\mu)}{{\rm BR}(h\to\tau\tau)},
\end{equation}
which is independent of the total width.
In particular, $R_{\tau\mu/\tau\tau}$ is insensitive
to the spectrum of the colored particles.
By combining $h\to\tau\tau$ and $h\to\tau\mu$ data we obtain the
experimentally allowed range for the ratio
of branching ratios
\begin{align}\label{eq:CurrentBoundHTauMu}
0.07\,(0.01) \lesssim R_{\tau\mu/\tau\tau}\lesssim 0.21\,(0.31) \qquad \text{at 68.3\% (95\%) \CL}
\end{align}
For this bound we assumed a parabolic $\chi^2$, {\it i.e.}\ gaussian errors, and
profiled over ${\rm BR}(h\to\tau\tau)$ to obtain the \CL\ interval on the ratio.
We thus focus on whether the MSSM can account for $R_{\tau\mu/\tau\tau} \gtrsim 0.1$.

The LHC measurements of the $h \to \tau\mu$ decay rate and their implications for new
physics have been discussed in the literature within various theoretical frameworks
\cite{Blankenburg:2012ex,Harnik:2012pb,Dery:2013rta,Dery:2014kxa,Arhrib:2012mg,Arroyo:2013tna,Celis:2013xja,Falkowski:2013jya,Campos:2014zaa,Sierra:2014nqa,Heeck:2014qea,Crivellin:2015mga,Dorsner:2015mja,Das:2015zwa,Bishara:2015cha,Bhattacherjee:2015sia,He:2015rqa,Altmannshofer:2015esa,Cheung:2015yga,Arganda:2015naa,Botella:2015hoa,Baek:2015mea}.
In particular, previous studies of $h\to\tau\mu$ within the supersymmetric framework have been carried out in Refs.~\cite{Brignole:2003iv,Brignole:2004ah,Arganda:2004bz,Arana-Catania:2013xma}.
In these studies the emphasis was on identifying the range of ${\rm BR}(h\to\tau\mu)$
that corresponds to generic points in the parameter space.
Indeed, we confirm that for generic supersymmetric parameters, ${\rm BR}(h\to\tau\mu)$ is
several orders of magnitude below the present experimental sensitivity, as found in these
previous works. We, however, are interested to learn whether, if ${\rm BR}(h\to\tau\mu)={\cal O}(0.01)$ is established
at the LHC, the MSSM will be not just disfavored but actually excluded.
To answer this question, we allow the parameters to be highly fine-tuned and far from generic.

The structure of this paper is as follows.
In Section \ref{sec:Requirements} we present our theoretical framework.
In Sections \ref{sec:ae} and \ref{sec:mu} we obtain the largest possible
$R_{\tau\mu/\tau\tau}$ that can arise from LFV from the $A$-terms and the slepton masses-squared, respectively, taking into account bounds from $\tau\to\mu\gamma$ and $h\to\gamma\gamma$ and from perturbativity.
In Section \ref{sec:ccb} we require, in addition,  that the electroweak symmetry
breaking minimum is the global one and, in particular, that there is no deeper minimum that is charge breaking.
We conclude in Section~\ref{sec:conclusions}.
Supplementary material is delegated to the Appendices.

\section{The theoretical framework}\label{sec:Requirements}
We consider the minimal supersymmetric SM.
We assume R-parity conservation (RPC), but make no assumptions, such as universality or alignment,
about the supersymmetric mass spectrum and mixing pattern.
Examining the MSSM in view of the three points emphasized above, we make the following observations:
\begin{itemize}
\item Lepton flavor is not an accidental symmetry of the MSSM.
\item Within the MSSM, FCNCs are always loop mediated.
\item The R-parity even scalar sector of the MSSM is a two Higgs doublet model (2HDM) with NFC type-II.
\end{itemize}
We now elaborate on each of these three features of the MSSM.

The supersymmetric part of the MSSM Lagrangian is minimally flavor violating:
the only supersymmetric sources of flavor violation are the Yukawa matrices of the SM.
Therefore, this part of the Lagrangian has the same accidental $U(1)_e\times U(1)_\mu\times U(1)_\tau$
symmetry as the SM. However, this is in general not the case for the soft supersymmetry breaking terms.
They have three sources of LFV:
\begin{equation}\label{eq:SleptonSoftMassMatrices}
\mathcal{L}_{\rm MSSM}^{\rm LFV} = -\tilde{m}^2_{L_{ij}} \tilde{L}^{\dagger}_i \tilde{L}_j -\tilde{m}^2_{R_{ij}} \tilde{\bar{E}}^{\dagger}_{i} \tilde{\bar{E}}_{j} - (A_{ij}^E h_d  \tilde{L}_i\tilde{\bar{E}}_{j} + {\rm h.c.})\,.
\end{equation}
Here $\tilde L_i$ are the $SU(2)$-doublet sleptons, $\tilde{\bar{E}}_{i}$ are the $SU(2)$-singlet charged sleptons, and $h_d$ is the $Y=-1/2$ Higgs doublet; $\tilde{m}^2_{L}$ is the $3\times3$ mass-squared matrix for the doublet sleptons, $\tilde{m}^2_{R}$ is the $3\times3$ mass-squared matrix for the singlet sleptons, and $A^E$ is the $3\times3$ matrix of trilinear scalar couplings.
Throughout this work we follow the conventions of Ref.~\cite{Rosiek:1995kg}.

Since either $h\to\tau\mu$ or $h\to\tau e$, but not both, can be large \cite{Blankenburg:2012ex},
we decouple in what follows the selectron, and consider only the $2\times2$ $\mu-\tau$ block
of each of the three matrices.

In addition to $\mathcal{L}_{\rm MSSM}^{\rm LFV}$, the following superpotential terms, involving the Higgs ($H_u,H_d$) and lepton ($L,\bar E$) superfields, are relevant to our study:
\begin{equation}
W_{H,L,E}=\mu H_u H_d+Y^E H_d L\bar E\,.
\end{equation}
In the charged lepton mass basis, $Y^E={\rm diag}(y_e,y_\mu,y_\tau)$.

Two additional parameters that affect our results are the angles $\beta$ and $\alpha$.
At the tree level they are traditionally defined as $\tan\beta=v_u/v_d$
(where $v_{u,d}=\langle H_{u,d}\rangle$), and $\alpha$ the rotation angle
from the $(h_d,h_u)$ basis to the mass basis of the neutral CP-even Higgs
mass eigenstates $(h,H)$.
However, we consider loop corrections, where subtleties arise in the definitions
of these parameters.
The definition that we use (and is particularly convenient for our purposes)
is given in Eq.~\eqref{eq:defab}. In what follows, we use the notations $t_\phi\equiv\tan\phi$, $c_\phi\equiv\cos\phi$, and $s_\phi\equiv\sin\phi$ for the various angles that are relevant to our analysis.

\begin{figure}[t]%
\centering
\subfloat[][]{\includegraphics{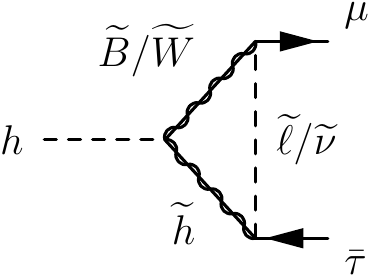}\label{fig:Fig1a}}%
\qquad
\subfloat[][]{\includegraphics{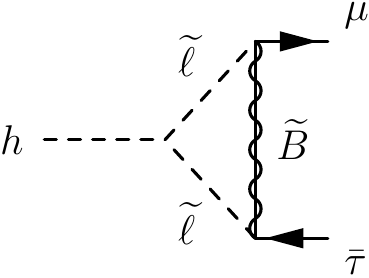}\label{fig:Fig1b}}%
\qquad
\subfloat[][]{\includegraphics{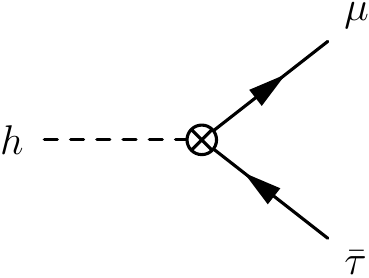}\label{fig:Fig1c}}%
\caption{Diagrams contributing to the one-loop amplitude for $h\to\tau\mu$.
$\otimes$ depicts the flavor off-diagonal counterterm from the field
renormalization $\delta Z_{\tau\mu}$.\label{fig:FeynmanDiagrams}}
\label{fig:cont}%
\end{figure}

Within the RPC MSSM, FCNC processes in general, and the $h\to\tau\mu$ decay in particular,
get no tree-level contributions.
The leading one-loop diagrams that contribute to this process are presented
in Fig.~\ref{fig:FeynmanDiagrams}.
In these diagrams, $\tilde\ell$ stands for the charged sleptons,
$\tilde\nu$ for the sneutrinos, $\tilde B$ for the bino and $\tilde W$ for the wino.

The diagrams of Fig.~\ref{fig:Fig1a} are proportional to
$y_\tau\times\sin2\theta\times\frac{\alpha}{4\pi}$, where $\theta$ is the smuon-stau
mixing angle.
In addition, this contribution is proportional to a loop function that depends on ratios
of sparticle mass parameters and is, at most, of ${\cal O}(1)$.
The electroweak loop factor of $\frac{\alpha}{4\pi}$ suppresses the amplitude by
three orders of magnitude with respect to the tree level $h\to\tau\tau$ decay.
Thus, these diagrams cannot generate $R_{\tau\mu/\tau\tau}\gsim0.1$, and we do not
consider them any further.

The diagrams of Fig.~\ref{fig:Fig1b} involve a trilinear scalar coupling.
We distinguish two cases:
\begin{enumerate}
\item The trilinear scalar coupling arises from the supersymmetric terms  $\mu Y^E$.
This case has two important features.
First, the source of LFV has to be either $(\tilde{m}^2_{L})_{\mu\tau}$ or $(\tilde{m}^2_{R})_{\mu\tau}$.
Second, the relevant Higgs field is $h_u$, while the tree level tau Yukawa coupling involves $h_d$.
In the limit of light 2HDM and heavy supersymmetry, the leading effect to $h\to\tau\mu$ arises from the
misalignment between the vacuum expectation value and the light mass eigenstate and is therefore proportional to $c_{\beta-\alpha}$.
Similarly to the diagrams of Fig.~\ref{fig:Fig1a}, this contribution to
$R_{\tau\mu/\tau\tau}$ is proportional to $[\sin2\theta\frac{\alpha}{4\pi}]^2$.
In this case, however, the contribution is proportional to the ratio of the dimensionful parameter
$\mu$ and the bino or slepton mass.
This factor can provide some enhancement.
\item The trilinear scalar coupling comes from the $A^E$ matrix.
Now, the source of LFV can be the trilinear coupling itself, namely $A^E_{\mu\tau}$ or $A^E_{\tau\mu}$.
Different from the previous case, the relevant Higgs field is $h_d$,
the same as the one that has the diagonal tree-level coupling $y_\tau$.
This contribution is, in general, not proportional to $y_\tau$.
Nevertheless, if the mass scale of the sleptons and/or the bino is somewhat heavier
than the electroweak scale, $m_{\rm SUSY}>v$, this contribution is suppressed by
$v^2/m_{\rm SUSY}^2$.
This decoupling behavior is clear because in this case in the limit of heavy SUSY
there is a single Higgs doublet so $h\to\tau\mu$ is mediated by the dimension-six
operator $\frac{\lambda_{ij}}{m_{\rm SUSY}^2} H^3 \bar{L}_i E_j$.
\end{enumerate}
In the next two sections we present how to maximize $h\to\tau\mu$ in each of those cases, taking into account relevant experimental bounds and perturbativity. The $A$-term case is analyzed in Section~\ref{sec:ae} and the slepton mass-squared case in Section~\ref{sec:mu}. The consequences of requiring that the global minimum is not charge breaking are analyzed, for both cases, in Section \ref{sec:ccb}.

The diagram of Fig.~\ref{fig:Fig1c} corresponds to the finite flavor off-diagonal
counterterm related to the field renormalization $\delta Z_{\tau\mu}$.
It ensures that lepton fields are canonically normalized and it is essential
to include it in order to have the correct decoupling behavior.
Its computation is outlined in Appendix~\ref{sec:ren}.

Finally, for the calculation of the ratio $R_{\tau\mu/\tau\tau}$, we also
need the tree-level contribution
\begin{equation}\label{eq:htt}
|{\cal M}(h\to\tau\tau)|^2= 2 m_h^2 \left(\frac{m_\tau}{v}\frac{s_\alpha}{c_\beta}\right)^2.
\end{equation}
Above we neglected terms that are suppressed by additional powers of external
fermion masses.
Our choice for defining $\alpha$ and $\beta$ at higher orders
in perturbation theory, given in Appendix \ref{app:ab},  ensures that Eq.~\eqref{eq:htt} remains valid also at the loop level.

\section{\texorpdfstring{LFV from the $\boldsymbol{A^E}$ terms}{%
                         LFV from the AE terms}\label{sec:ae}}

\begin{figure}[t]
\begin{center}
\includegraphics[]{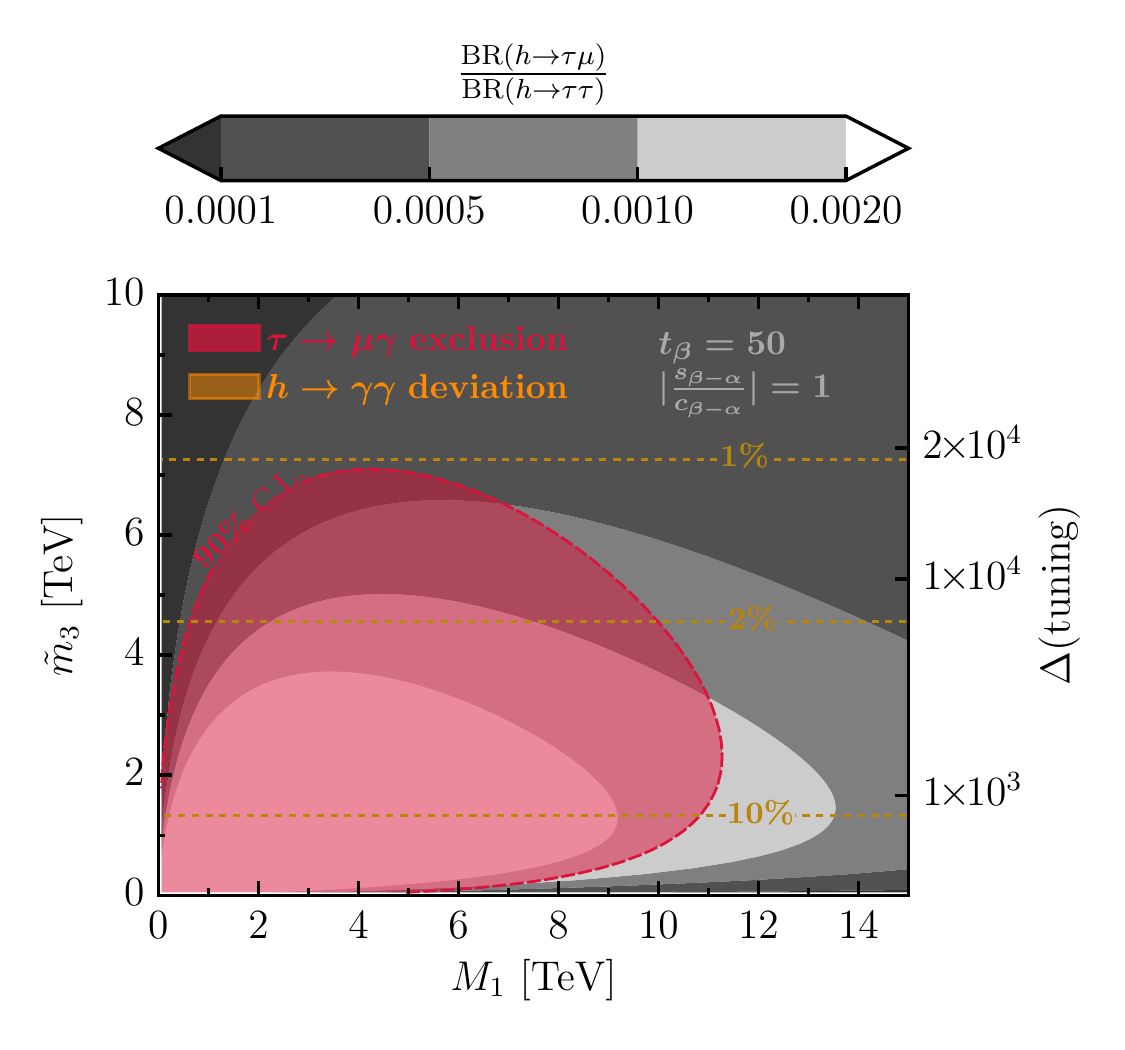}
\end{center}
\caption{Contours of $R_{\tau\mu/\tau\tau}$ in the $M_1-\tilde m_3$ for the case
of LFV from the A-term $A_{\mu\tau}$. The value of the A-term is the maximal
allowed by perturbativity and satisfies vacuum stability constraints;
the associated tuning is indicated on the right y-axis.
The red region in dashed lines is excluded by the bound on $\tau\to\mu\gamma$.
Orange horizontal lines indicate the deviation of the partial $h\to\gamma\gamma$
width with respect to the SM value.
}
\label{fig:M1m3aterms}
\end{figure}

Consider the case that the $\tilde{m}^2_{L}$ and $\tilde{m}^2_{R}$ matrices are diagonal,
and the only source of LFV is the $A^E$ matrix.
Both $A_{\mu\tau}^E$ and $A_{\mu\tau}^E$ contribute to the $h\to\tau\mu$ decay.
The analysis is simplified if the $\tilde\tau_L^\dagger\tilde\tau_R$ entry in the
mass-squared matrix can be neglected.
Then, the $4\times4$ mass-squared matrix decomposes into two $2\times2$ blocks.
For concreteness, we analyze the $\tilde\mu_L-\tilde\tau_R$ block.
The analysis of the $\tilde\tau_L-\tilde\mu_R$ block is similar.
The relevant part of the slepton mass-squared matrix has the following form:
\begin{align}
\tilde{\mathcal{M}}^2 =
\begin{pmatrix}
\tilde{m}^2_{\mu_L}              & \frac{v_d A_{\mu\tau}}{\sqrt{2}} \\
\frac{v_d A_{\mu\tau}}{\sqrt{2}} & \tilde{m}^2_{\tau_R}
\end{pmatrix}
\end{align}
The mixing angle $\theta$, which rotates from the interaction basis $(\tilde\mu_L,\tilde\tau_R)$ to the mass basis $(\tilde\ell_3,\tilde\ell_2)$, is given by
\begin{equation}
\tan2\theta=\frac{\sqrt2v_d A_{\mu\tau}}{\tilde m_{\tau_R}^2-\tilde m_{\mu_L}^2}.
\end{equation}
In the mass basis, the trilinear scalar couplings are given by
\begin{equation}
\mathcal{L}_{h\tilde\ell\tilde\ell}
 = \frac{A_{\mu\tau} s_\alpha}{\sqrt{2}} h \left[s_{2\theta} \left(\tilde{\ell}^\ast_2 \tilde{\ell}_2 - \tilde{\ell}^\ast_3\tilde{\ell}_3\right)+ c_{2\theta}\left(\tilde{\ell}^\ast_2 \tilde{\ell}_3 + \tilde{\ell}^\ast_3 \tilde{\ell}_2\right)\right]
\end{equation}
and the bino-lepton-slepton couplings by
\begin{equation}
\mathcal{L}_{\tilde B\tilde\ell\ell} =  \frac{g'}{\sqrt{2}} \tilde{B} \left(s_\theta \tilde{\ell}^\ast_2 + c_\theta \tilde{\ell}^\ast_3\right) \mu_L - g' \sqrt{2} \tilde{B} \left(c_\theta \tilde{\ell}_2 - s_\theta \tilde{\ell}_3\right) \bar{\tau}_R + {\rm h.c.}
\end{equation}

To zeroth order in the expansion of external momenta over SUSY masses, we obtain the
following contributions to the decay amplitude of $h \to \tau^+\mu^-$:
\begin{equation}\label{eq:SoFinally}
\begin{split}
{\rm Fig.~\ref{fig:Fig1c}}= &  -i \frac{\alpha M_1 s_{2\theta} s_\alpha}{8  \pi c^2_{W} v c_\beta}  \left(x_2-x_3\right) I_3 (1,x_2,x_3)~P_R,\\
{\rm Fig.~\ref{fig:Fig1b}} = &i  \frac{ \alpha  A_{\mu\tau} s_\alpha}{8\sqrt{2} \pi c^2_{W} M_1}  \left\{
s^2_{2\theta} \left[I_3(1,x_2,x_2)+ I_3(1,x_3,x_3) \right]
+2c^2_{2\theta} I_3(1,x_2,x_3)
\right\}~P_R,
\end{split}
\end{equation}
with $c_W^2\equiv\cos^2\theta_W$, $P_R=(1+\gamma_5)/2$, $x_i\equiv\tilde m_i^2/M_1^2$, and
\begin{equation}
I_3(x,y,z)=\frac{xy\log(x/y)+yz\log(y/z)+zx\log(z/x)}{(x-y)(y-z)(z-x)}.
\end{equation}
For brevity we suppressed in Eq.~\eqref{eq:SoFinally} the spinors for the lepton fields.
The amplitude for $h \to {\tau^-}\mu^+$ is the same as the one in
Eq.~\eqref{eq:SoFinally} after exchanging $P_R$ with $P_L=(1-\gamma_5)/2$.

The sum of the two amplitudes then reads:
\begin{multline}
\mathcal{M}(h\to \tau^+\mu^-) = i \frac{\alpha A_{\mu\tau} s_\alpha}{8\sqrt2\pi c^2_{W}M_1}
\biggl\{s^2_{2\theta}\left[ I_3(1,x_2,x_2) + I_3(1,x_3,x_3)\right]+\\
+2\left[c_{2\theta}^2-s_{2\theta}\frac{M_1^2(x_2-x_3)}{\sqrt2vc_\beta A_{\mu\tau}}\right]I_3 (1,x_2,x_3)\biggl\}~P_R.
\end{multline}

We see that in the limit of small mixing, $s_{2\theta}\ll1$
($vc_\beta A_{\mu\tau}\ll\tilde m_2^2-\tilde m_3^2$), we have
${\cal M}(h\to\tau\mu)={\cal O}(s_{2\theta}^2)={\cal O}(v^2/m_{\rm SUSY}^2)$,
as argued in the previous section.
To estimate the largest possible value of $R_{\tau\mu/\tau\tau}$, we take, however,
the limit of maximal mixing, $\sin^22\theta=1$.
This limit is obtained by fine-tuning $\tilde m_{\mu_L}^2=\tilde m_{\tau_R}^2$ and,
consequently, $\tilde m_2^2-\tilde m_3^2=\sqrt{2}vc_\beta A_{\mu\tau}$.
In this case the sum of $|{\cal M}(h\to\tau^+\mu^-)|^2$ and $|{\cal
M}(h\to\tau^-\mu^+)|^2$ reads
\begin{equation}
\begin{split}
|\mathcal{M}(h\to \tau\mu)|^2 & = m_h^2 \frac{\alpha^2 s^2_\alpha}{64\pi^2 c^4_{W}} \frac{A_{\mu\tau}^2}{M_1^2}
\left[ I_3(1,x_2,x_2) + I_3(1,x_3,x_3) -2 I_3 (1,x_2,x_3) \right]^2.
\end{split}
\end{equation}
To check that our result has the correct decoupling behavior for $m_{\rm SUSY}\gg v$, we evaluate the terms in parenthesis in the limit $(\tilde m_2^2-\tilde m_3^2)\ll  (\tilde m_2^2+\tilde m_3^2), M_1^2$:
\begin{equation}
\vert\mathcal{M}(h\to \tau\mu)\vert^2=
m_h^2\frac{\alpha^2 s_\alpha^2}{144\pi^2 c^4_{W}}\frac{A_{\mu\tau}^2}{M_1^2}
\left(\frac{x_2-x_3}{x_2+x_3}\right)^4
\frac{\left(1-6x_{23}+3x_{23}^2+2x_{23}^3-6x_{23}^2\log x_{23}\right)^2}
{(1-x_{23})^8},
\end{equation}
where $x_{23}\equiv(\tilde m_2^2+\tilde m_3^2)/(2M_1^2)$.
The amplitude indeed scales as  $\frac{(x_2-x_3)^2}{(x_2+x_3)^2} \sim  v^2/m_{\rm SUSY}^2$.

The amplitude grows as $A_{\mu\tau}$. There is, however, a perturbativity bound on $A_{\mu\tau}$:
\begin{equation}\label{eq:atmper}
\left\vert \frac{A_{\mu\tau} s_\alpha}{\sqrt{2}} \right\vert\lesssim 4\pi \tilde{m}_3\,.
\end{equation}
The fine tuning in models with $m_h\ll A_{\mu\tau}$ is of order \cite{Papucci:2011wy}
\begin{equation}
\Delta \simeq \frac{1}{16\pi^2}\frac{A^2_{\mu\tau}}{m_h^2}\,.
\label{}
\end{equation}
We denote the ratio $R_{\tau\mu/\tau\tau}$ that corresponds to maximal mixing, $s^2_{2\theta}=1$,
and maximal perturbative $A_{\mu\tau}=4\sqrt2\pi \tilde{m}_3/s_\alpha$
by $R_{\tau\mu/\tau\tau}^{\rm max}$:
\begin{equation}
R_{\tau\mu/\tau\tau}^{\rm max}=\left\{
\frac{\alpha}{2 c^2_W}\frac{v}{ m_\tau} \frac{c_\beta}{s_\alpha}\frac{\tilde m_3}{M_1}
\left[ I_3(1,x_2,x_2) + I_3(1,x_3,x_3) -2 I_3 (1,x_2,x_3) \right]\right\}^2.
\end{equation}

In Fig.~\ref{fig:M1m3aterms}
we show the value of $R_{\tau\mu/\tau\tau}^{\rm max}$
in the $\tilde m_3-M_1$ plane.
Here, $A_{\mu\tau}=4\sqrt2\pi \tilde{m}_3/|s_\alpha|$,
$\tilde m_2^2=\tilde m_3^2+\sqrt{2}vc_\beta A_{\mu\tau}$, and $|s_\alpha /c_\beta| = 1$. We emphasize that  $|s_\alpha /c_\beta| = 0.75$ gives an $\mathcal{O}(1)$ enhancement but is forbidden by vacuum stability, which will be discussed in more detail below.
Also depicted in this plot is the region excluded by the upper bound on $\tau\to\mu\gamma$
(for details see Appendix \ref{app:tmg}) and contours of deviation of the
$h\to\gamma\gamma$ partial width with respect to the SM one (see Appendix~\ref{app:hgg}).

We conclude that, with $A_{\mu\tau}$ being the source of LFV, we have
\begin{equation}\label{eq:perarub}
R_{\tau\mu/\tau\tau}\lsim0.0015,
\end{equation}
below the near-future sensitivity of ATLAS and CMS.
Even by including both $A_{\mu\tau}$ and $A_{\tau\mu}$ at the same time,
$R_{\tau\mu/\tau\tau}\lsim0.002$.

\section{\texorpdfstring{LFV from the $\boldsymbol{\tilde{m}^2_{L}}$ terms}{%
                         LFV from the m2L terms}\label{sec:mu}}

Consider the case that the sources of LFV are the matrices
$\tilde{m}^2_{L}$ and $\tilde{m}^2_{R}$.
To obtain $R_{\tau\mu/\tau\tau}$ as large as ${\cal O}(0.1)$, there must be no
additional suppression from the mixing angle or from the loop function.
At least one of $\tilde{m}^2_{L}$ and $\tilde{m}^2_{R}$
has to be anarchic in the $\mu-\tau$ sector to have a mixing angle
of order one.
While large $\tilde\mu-\tilde\tau$ mixing is a necessary condition,
if it is large in both $\tilde{m}^2_{L}$ and $\tilde{m}^2_{R}$,
the $\tau$-lepton in Fig.~\ref{fig:Fig1b} can be replaced with a muon,
which implies that ${\rm BR}(h\to\mu\mu)\sim{\rm BR}(h\to\tau\mu)$.
Given our requirement that $R_{\tau\mu/\tau\tau}\gsim0.1$, and the experimental
upper bound  on ${\rm BR}(h\to\mu\mu)$ \cite{Khachatryan:2014aep,Aad:2014xva}, this case is disfavored.
Thus, either $\tilde{m}^2_{L}$ or $\tilde{m}^2_{R}$ has to be near-diagonal.
For concreteness, we take $\tilde{m}^2_{L}$ to be anarchic and $\tilde{m}^2_{R}$
to be diagonal.
Hence, we focus on the $3\times3$ block of
$(\tilde\mu_L,\tilde\tau_L,\tilde\tau_R)$ in the slepton mass-squared matrix.

The relevant part of the slepton mass-squared matrix has the form:
\begin{align*}
\tilde{\mathcal{M}}^2 =
\begin{pmatrix}
(\tilde{m}^2_L)_{\mu\mu} & (\tilde{m}^2_L)_{\mu\tau} & 0 \\
(\tilde{m}^2_L)_{\mu\tau}^* & (\tilde{m}^2_L)_{\tau\tau} & -m_\tau\mu t_\beta \\
0 & -m_\tau\mu t_\beta &(\tilde{m}^2_{R})_{\tau\tau}
\end{pmatrix},
\end{align*}
where, for simplicity, we set $A^E=0$ and $y_\mu=0$.

We denote by $\tilde U$ the mixing matrix that rotates from the
interaction basis $(\tilde\mu_L,\tilde\tau_L,\tilde\tau_R)$
to the mass basis $(\tilde\ell_1,\tilde\ell_2,\tilde\ell_3)$.
To maximize the rate of $h\to\tau\mu$, it is best if the dominant contribution
comes from the lightest slepton mass eigenstate, $\tilde\ell_3$.
The mixing angles that enter the amplitude are
\begin{equation}\label{eq:max33}
\tilde U_{3\mu_L}^*\tilde U_{3\tau_R}\times2{\cal R}e(\tilde U_{3\tau_L}\tilde U_{3\tau_R}^*).
\end{equation}
We are interested in estimating the largest possible contribution to $h\to\tau\mu$.
Therefore, we are interested in the values of $\tilde U_{3\alpha}$ that maximize
Eq.~\eqref{eq:max33}:
\begin{equation}\label{eq:bestU}
\tilde U_{3\alpha}=(1/2,1/2,1/\sqrt2).
\end{equation}
The way to achieve Eq.~\eqref{eq:bestU} is by two tunings of entries of $\tilde{\mathcal{M}}^2$.
First, we set  $(\tilde{m}^2_L)_{\mu\mu}=(\tilde{m}^2_L)_{\tau\tau}$.
Then, we extract the two eigenvalues of the $\tilde m^2_L$ matrix.
We take the heavier eigenvalue $(\tilde m^2_L)_+$ to be very large, so that the
corresponding mass eigenstate $\tilde\ell_+=\frac{1}{\sqrt2}(\tilde\tau_L+\tilde\mu_L)$
decouples. We are left with an effective two-slepton framework,
$\tilde\ell_-=\frac{1}{\sqrt2}(\tilde\tau_L-\tilde\mu_L)$ and $\tilde\tau_R$:
\begin{align*}
\tilde{\mathcal{M}}^2 =
\begin{pmatrix}
(\tilde m_L^2)_- 		& -m_\tau\mu t_\beta/\sqrt{2} \\
-m_\tau\mu t_\beta/\sqrt{2} 	& (\tilde{m}^2_{R})_{\tau\tau}
\end{pmatrix},
\end{align*}
where $(\tilde m_L^2)_-=(\tilde{m}^2_L)_{\tau\tau}-(\tilde{m}^2_L)_{\mu\tau}$.
The mixing angle $\theta$, which rotates from the interaction basis
$(\tilde\ell_-,\tilde\tau_R)$ to the mass basis $(\tilde\ell_3,\tilde\ell_2)$,
is given by
\begin{equation}
\tan2\theta=\frac{\sqrt{2}m_\tau\mu t_\beta}{(\tilde m_L^2)_- - (\tilde m_R^2)_{\tau\tau}}.
\end{equation}
The trilinear scalar couplings in the mass basis are given by
\begin{equation}
\mathcal{L}_{h\tilde\ell\tilde\ell}
 = -\frac{m_{\tau}\mu c_\alpha}{\sqrt2vc_\beta} h \left[s_{2\theta} \left(\tilde{\ell}^\ast_2 \tilde{\ell}_2 - \tilde{\ell}^\ast_3\tilde{\ell}_3\right)- c_{2\theta}\left(\tilde{\ell}^\ast_2 \tilde{\ell}_3 + \tilde{\ell}^\ast_3 \tilde{\ell}_2\right)\right]\,,
\end{equation}
and the bino--lepton--slepton couplings are given by
\begin{equation}
\mathcal{L}_{\tilde B\tilde\ell\ell} =  \frac{g'}{2} \tilde{B} \left(c_\theta \tilde{\ell}^\ast_2 + s_\theta \tilde{\ell}^\ast_3\right) (\tau_L-\mu_L) + \sqrt{2}g' \tilde{B} \left(s_\theta \tilde{\ell}_2 - c_\theta \tilde{\ell}_3\right) \bar{\tau}_R + {\rm h.c.}
\end{equation}
Given this Lagrangian we can compute the $h\to\tau\mu$ amplitude along lines similar to the analysis of the LFV $A$-terms.
There is the one-loop contribution (Fig.~\ref{fig:Fig1b}) and field renormalization
contribution (Fig.~\ref{fig:Fig1c}).

As a second step in obtaining optimal mixing, we tune $(\tilde m^2_L)_-=(\tilde{m}^2_{R})_{\tau\tau}$ to generate maximal $\tilde\ell_--\tilde\tau_R$ mixing.
This tunning fixes the mass difference of the two slepton mass eigenstates to be $\tilde{m}^2_2 - \tilde{m}^2_3 = \sqrt{2} m_\tau \mu t_\beta$.
We obtain, for the specific mixing pattern of Eq.~\eqref{eq:bestU}:
\begin{multline}
{\cal M}(h\to\tau^+\mu^-)=-i\frac{\alpha\mu m_\tau}{16\pi c_W^2 M_1 v}\left\{
c_{\beta-\alpha}\left[I_3(1,x_2,x_2)+I_3(1,x_3,x_3)
+2t_\beta^2I_3(1,x_2,x_3)\right]\right.\\
+\left.s_{\beta-\alpha}t_\beta\left[I_3(1,x_2,x_2)+I_3(1,x_3,x_3)
-2I_3(1,x_2,x_3)\right]\right\}P_R\,.
\end{multline}
We wrote this expression in a way that transparently exposes the correct decoupling limit.
We remind the reader that both $m_\tau/v$ and $\mu/M_1$ are finite in the decoupling limit.
Then, the first term scales like $c_{\beta-\alpha}$ and the second term scales like
$(\tilde m_2^2-\tilde m_3^2)^2/(\tilde m_2^2+\tilde m_3^2)^2$, both of which are
proportional to $v^2/m_{\rm SUSY}^2$.

For the ratio of branching ratios, we find:
\begin{equation}
R_{\tau\mu/\tau\tau}=\left(\frac{\alpha\mu}{16\pi c_W^2 M_1}\right)^2\left[2t_\beta I_3(1,x_2,x_3)-\frac{c_{\beta-\alpha}+s_{\beta-\alpha}t_\beta}{s_{\beta-\alpha}-c_{\beta-\alpha}t_\beta}
\sum_{i=2,3}I_3(1,x_i,x_i)\right]^2.
\end{equation}
The ratio grows as $(\mu/M_1)^2$. However, there is a perturbativity bound on $\mu$:
\begin{equation}\label{eq:muper}
\frac{m_\tau\mu(c_{\beta-\alpha}+s_{\beta-\alpha} t_\beta)}{\sqrt2 v}\lsim4\pi\tilde m_3\,.
\end{equation}
The fine tuning in models with $m_h\ll\mu$ is of order \cite{Papucci:2011wy}
\begin{equation}
\Delta \simeq \frac{2\mu^2}{m_h^2}\,.
\label{}
\end{equation}
We denote the ratio $R_{\tau\mu/\tau\tau}$ which corresponds to
$\tilde U_{3\alpha}=(1/2,1/2,1/\sqrt2)$ and to $\mu$ at the perturbative bound of
Eq.~\eqref{eq:muper} by $R_{\tau\mu/\tau\tau}^{\rm max}$:
\begin{equation}\label{eq:RmaxMuTerm}
R_{\tau\mu/\tau\tau}^{\rm max}=\left\{\frac{\alpha v \sqrt{x_3}}{2 \sqrt{2} m_\tau c_W^2}
\left[\frac{2I_3(1,x_2,x_3)}{s_{\beta-\alpha}+c_{\beta-\alpha}/t_\beta}
-\frac{\sum_{i=2,3}I_3(1,x_i,x_i)}{s_{\beta-\alpha}-c_{\beta-\alpha}t_\beta}
\right]\right\}^2.
\end{equation}

The parameters $c_{\beta-\alpha}$ and $t_\beta$ play a crucial role on the value of
$R_{\tau\mu/\tau\tau}^{\rm max}$.
The allowed range in the $c_{\beta-\alpha}-t_\beta$ plane is shown in Fig.~\ref{fig:cos_ba_t_b} of
Appendix~\ref{app:ab}.
The upper bound on $R_{\tau\mu/\tau\tau}^{\rm max}$ is different for the bulk
region and the peninsula region.
The peninsula region corresponds to the parameter space in which the $hVV$ and $h\gamma\gamma$
couplings are close to their SM values, while the $h\tau\tau$ coupling has the same absolute
value but opposite sign.

It is interesting to note that, for $M_1,\tilde m_3\gg v$, the sleptons are
quasi-degenerate and Eq.~\eqref{eq:RmaxMuTerm} takes the form
\begin{equation}
R_{\tau\mu/\tau\tau}^{\rm max}=\left\{\frac{\alpha v}{\sqrt{2} m_\tau c_W^2}
\sqrt{x_3}I_3(1,x_3,x_3)
\left[\frac{c_{\beta-\alpha}t_\beta}{s_{\beta-\alpha}(s_{\beta-\alpha}-c_{\beta-\alpha} t_\beta)}
\right]\right\}^2.
\end{equation}
The maximum of the loop function is for $\tilde m_3/M_1\approx0.47$ independently
of their individual values.
The best fit point for the trigonometric factor depends on whether we are in the
bulk or in the peninsula regions.

In the right panels of Figs.~\ref{fig:M1m3mu} and ~\ref{fig:M1m3mubranch} we show the value of $R_{\tau\mu/\tau\tau}^{\rm max}$ in the $\tilde m_3-M_1$ plane. Here, $m_\tau\mu=4\sqrt2\pi \tilde{m}_3 v/(c_{\beta-\alpha}+s_{\beta-\alpha}t_\beta)$,
and $\tilde m_2^2=\tilde m_3^2+\sqrt{2} m_\tau\mu t_\beta$.
Also depicted in these plots is the region excluded by the upper bound on
$\tau\to\mu\gamma$ (see Appendix \ref{app:tmg}) and the deviation of the partial width of
$h\to\gamma\gamma$ with respect to the SM (see Appendix \ref{app:hgg}).

Fig.~\ref{fig:M1m3mu} corresponds to $c_{\beta-\alpha}$ and $t_\beta$ in the bulk region.
We conclude that, with $(\tilde m^2_L)_{\mu\tau}$ being the source of LFV while
also being in the bulk region
\begin{equation}\label{eq:perboubul}
R_{\tau\mu/\tau\tau}\lsim0.035 \qquad \text{for } \left|c_{\beta-\alpha} t_\beta\right| \ll 1,
\end{equation}
below the near-future sensitivity of ATLAS and CMS.

Fig.~\ref{fig:M1m3mubranch} corresponds to $c_{\beta-\alpha}$ and $t_\beta$ in the peninsula region.
Here, much higher values of $R_{\tau\mu/\tau\tau}$ can be reached. In particular, the present upper bound
on this ratio (Eq.~\eqref{eq:CurrentBoundHTauMu}),
\begin{equation}\label{eq:satub}
R_{\tau\mu/\tau\tau} \lsim 0.31\qquad \text{for } c_{\beta-\alpha} t_\beta \simeq 2,
\end{equation}
can be saturated.

\section{Charge breaking minima}\label{sec:ccb}
\begin{figure}[t]%
\centering
\subfloat[][]{\includegraphics[scale=0.7]{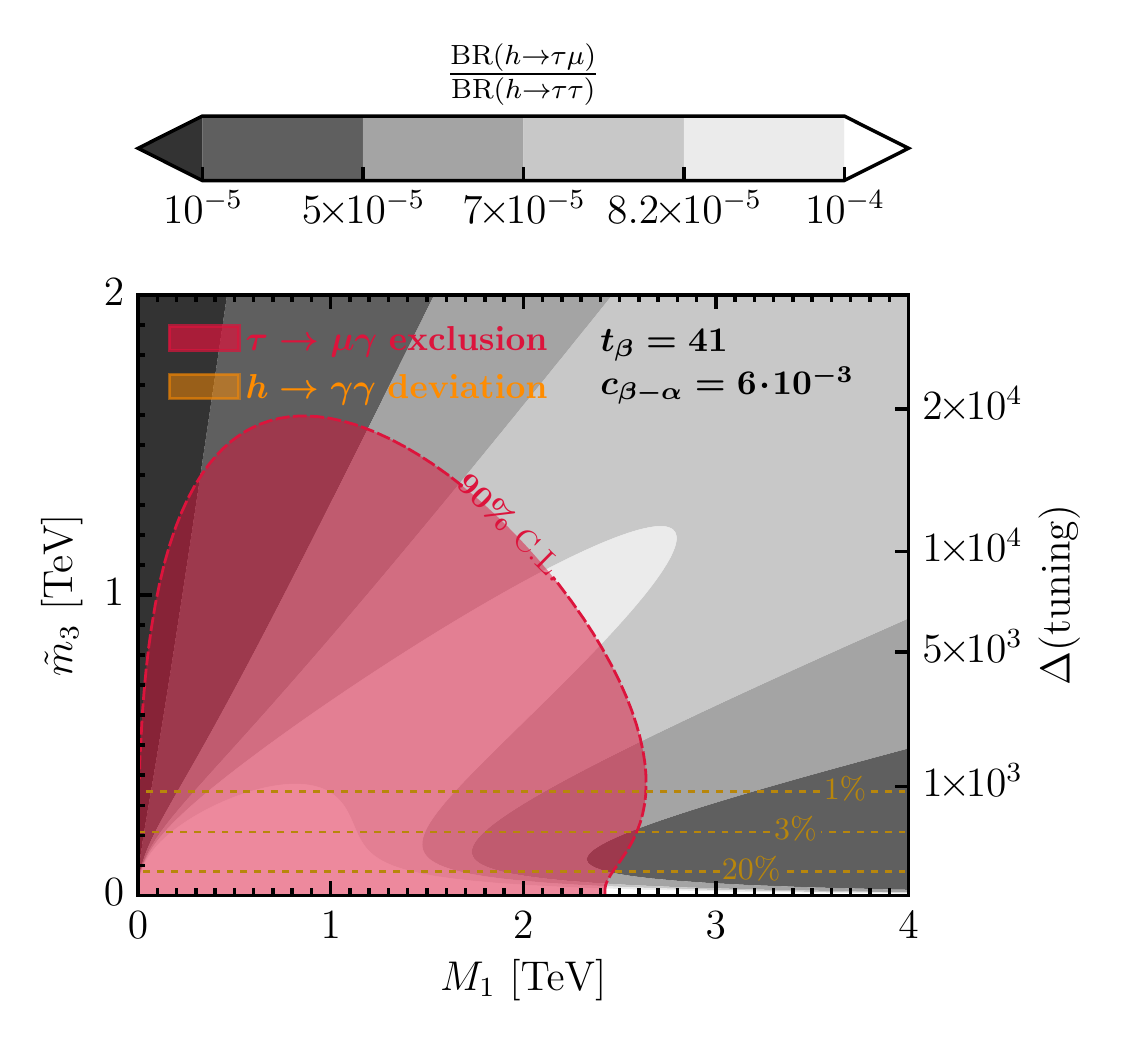}}%
\subfloat[][]{\includegraphics[scale=0.7]{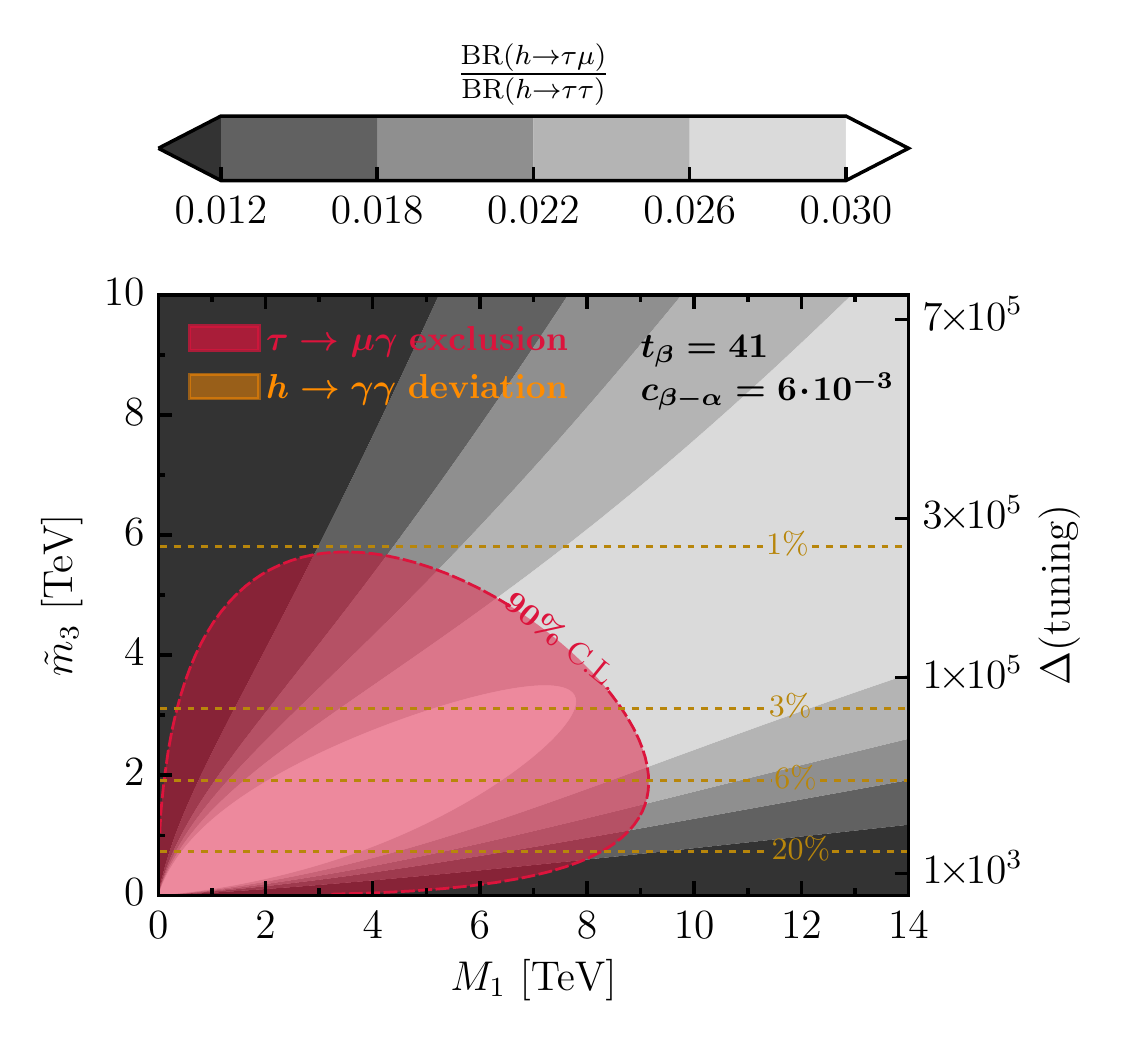}}%
\caption{Contours of $R_{\tau\mu/\tau\tau}$ in the $M_1-\tilde m_3$ for the case
of LFV from the slepton mass-squared term $(\tilde m^2_L)_{\mu\tau}$.
For $\cos(\beta-\alpha)$ and $\tan\beta$ we take
the values in the bulk region of Fig.~\ref{fig:cos_ba_t_b} that maximize $R_{\tau\mu/\tau\tau}$.
In the left (right) panel the value of the $\mu$ is the maximal
allowed by vacuum stability (perturbativity); the associated tuning is indicated on the right y-axis.
The red region in dashed lines is excluded by the bound on $\tau\to\mu\gamma$.
Orange horizontal lines indicate the deviation of the partial $h\to\gamma\gamma$
width with respect to the SM value.}
\label{fig:M1m3mu}
\end{figure}
\begin{figure}[t]%
\centering
\subfloat[][]{\includegraphics[scale=0.7]{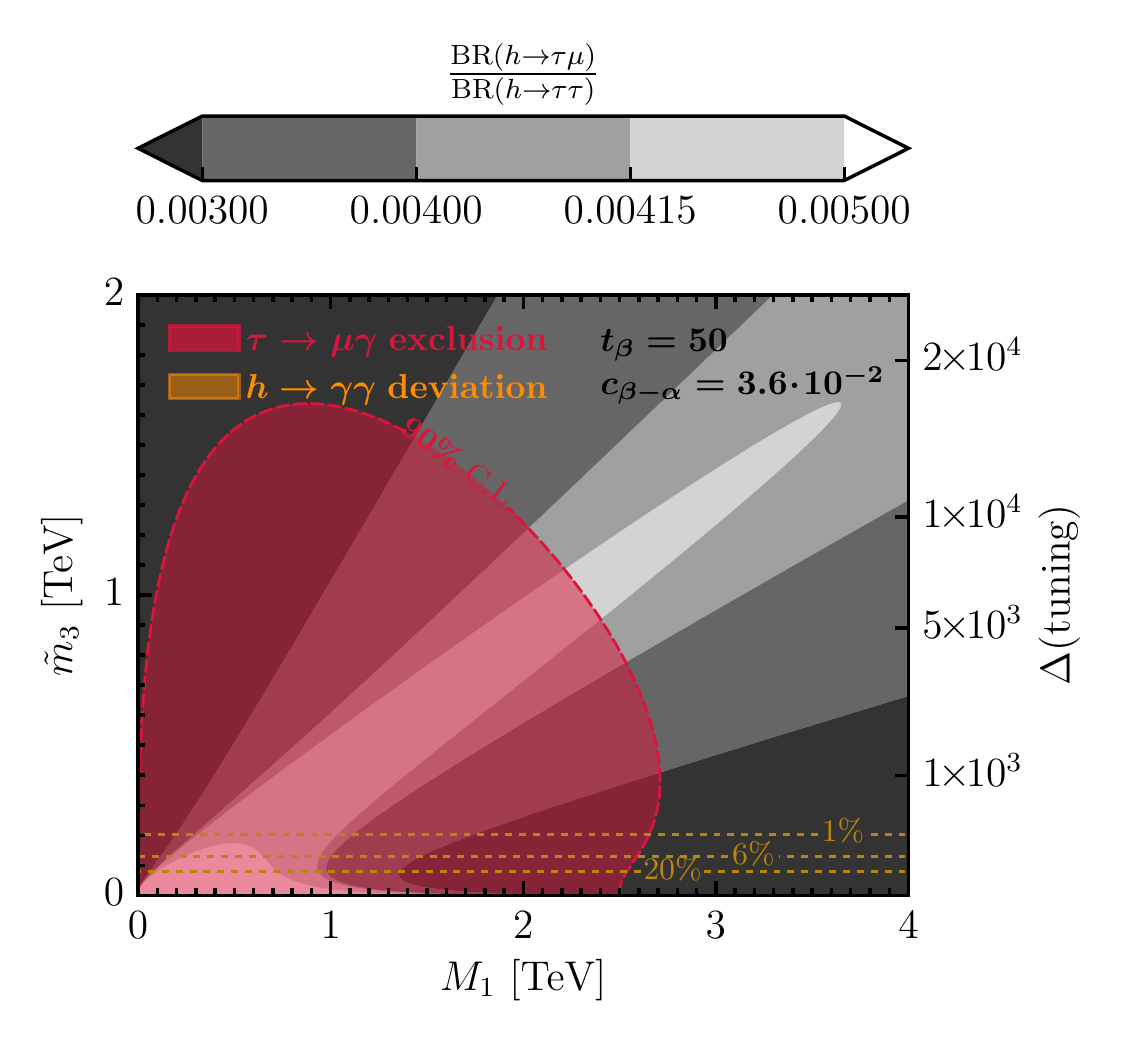}}%
\subfloat[][]{\includegraphics[scale=0.7]{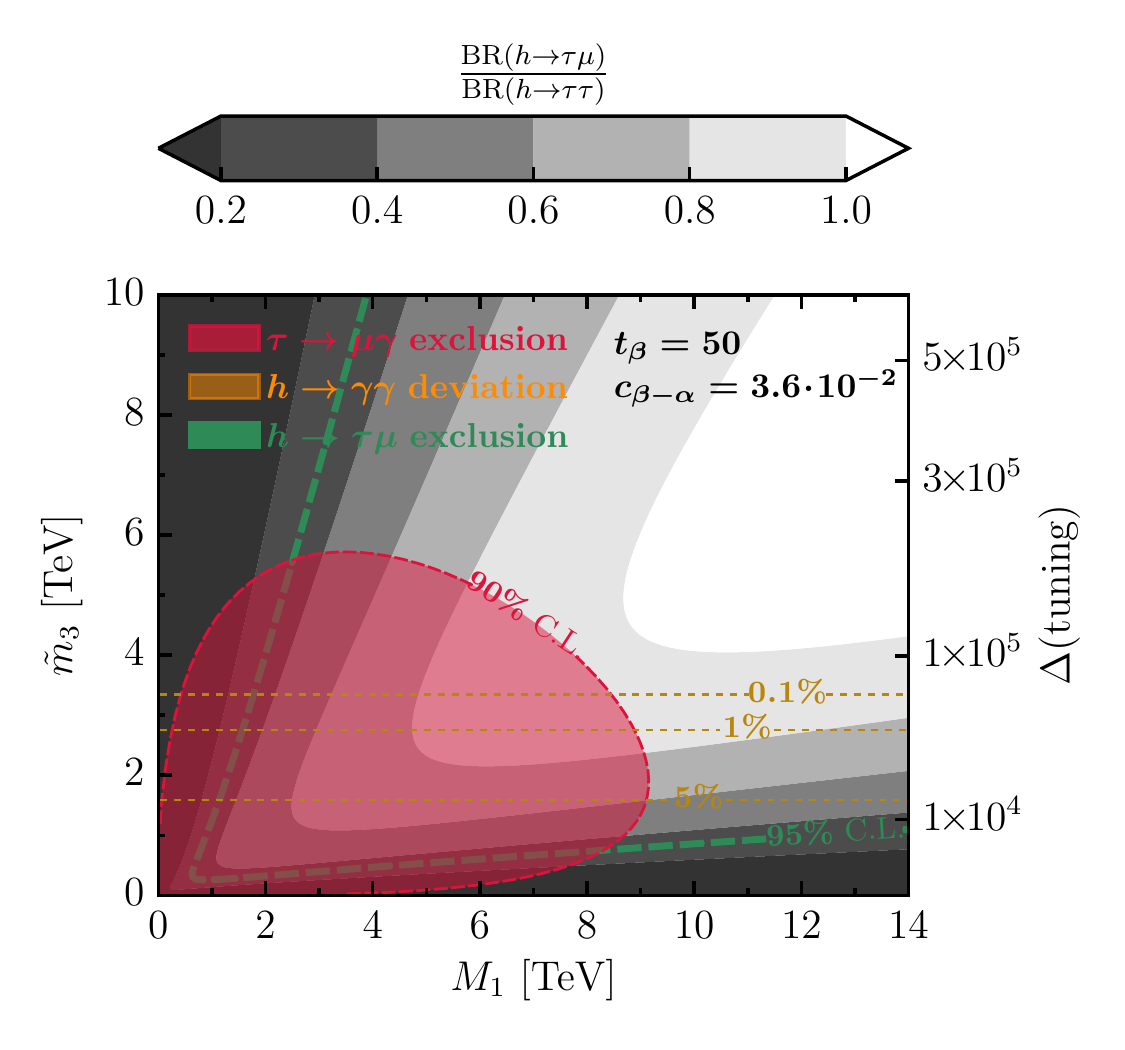}}%
\caption{Contours of $R_{\tau\mu/\tau\tau}$ in the $M_1-\tilde m_3$ for the case
of LFV from the slepton mass-squared term $(\tilde m^2_L)_{\mu\tau}$.
For $\cos(\beta-\alpha)$ and $\tan\beta$ we take
the values in the peninsula region of Fig.~\ref{fig:cos_ba_t_b} that maximize $R_{\tau\mu/\tau\tau}$.
In the left (right) panel the value of the $\mu$ is the maximal
allowed by vacuum stability (perturbativity); the associated tuning is indicated on the right y-axis.
The red region in dashed lines is excluded by the bound on $\tau\to\mu\gamma$.
Orange horizontal lines indicate the deviation of the partial $h\to\gamma\gamma$
width with respect to the SM value. The green dashed line indicates the direct upper bound on $R_{\tau\mu/\tau\tau}$.}
\label{fig:M1m3mubranch}
\end{figure}
In previous sections we established that large trilinear scalar couplings enable, in principle, enhancement of the MSSM contributions to $h\to\tau\mu$ well above the estimate from naive dimensional analysis. Such large trilinear couplings might lead, however, to charge breaking minima that are lower than the electroweak symmetry breaking one. In this section we obtain upper bounds on the trilinear couplings by requiring that the global minimum is not charge breaking.

We are particularly concerned whether the corner of parameter space, in which the MSSM
saturates the upper bound (see Eq.~\eqref{eq:satub}) while being consistent with experimental
constraints and the perturbativity bound, does not lead to a global minimum that is charge breaking.
The relevant scalar fields are the Higgs and slepton fields.
Similar bounds on $\mu$ have been investigated in the past
\cite{Rattazzi:1995gk,Hisano:2010re,Altmannshofer:2012ks,Carena:2012mw,Altmannshofer:2014qha}.
In our case, the dangerous directions in the field space are those where there is no dependence on
the heavy slepton mass-squared eigenvalue $\tilde m^2_{L_+}$.
Thus, we focus on the following direction:
\begin{equation}
\langle H_u^0\rangle=\langle\tilde\tau_R\rangle=\langle\tilde\ell_{-}\rangle=f.
\end{equation}
In this direction, the soft breaking terms and F-terms are the following:
\begin{multline}
V_{F+{\rm soft}}=\tilde m^2_{L_-}\langle\tilde\ell^*_{-}\rangle\langle\tilde\ell_{-}\rangle
+\tilde m^2_{R}\langle\tilde\tau^*_{R}\rangle\langle\tilde\tau_{R}\rangle
+\frac{1}{\sqrt2}y_\tau\mu\left(\langle H_u\rangle\langle\tilde\tau^*_{R}\rangle\langle\tilde\ell_{-}\rangle+\text{h.c.}\right)\\
+\frac12 y_\tau^2\langle\tilde\tau^*_{R}\rangle\langle\tilde\tau_{R}\rangle\langle\tilde\ell^*_{-}\rangle\langle\tilde\ell_{-}\rangle,
\end{multline}
where we have used that, for $t_\beta\gg1$ and $c_{\beta-\alpha}\ll 1$,
$(M_{H_u}^2+ \vert \mu\vert^2) \ll \Lambda^2_{\rm SUSY}$ to neglect the
contribution from the $\langle H_u\rangle^2$ term.
Using our ansatz, $\tilde m^2_{L_-}=\tilde m^2_{R}=\frac12(\tilde m_2^2+\tilde m_3^2)$,
and adding the D-term, we obtain
\beq
V\simeq\left(\tilde m_2^2+\tilde m_3^2+\sqrt{2}y_\tau\mu f+\frac{y_\tau^2+g^2+g^{\prime2}}{2}f^2\right)f^2.
\eeq
We require that the minimum at $f=0$ should be deeper than minima with $f\neq0$. This is equivalent to requiring that the discriminant for the term in parenthesis is negative:
\beq
\mu^2\leq (\tilde m_2^2+\tilde m_3^2)\frac{y_\tau^2+g^2+g^{\prime2}}{y_\tau^2}.
\eeq
Using the relation that follows from our ansatz, $\tilde m_2^2=\tilde m_3^2+\sqrt2 m_\tau\mu t_\beta$, we find, to leading order in $v/\tilde m_3$,
\beq\label{eq:mustability}
|\mu|\lsim\sqrt{\frac{2}{y_\tau^2}(y_\tau^2+g^2+g^{\prime2})}\tilde m_3\sim 2.5 \tilde m_3,
\eeq
where in the last equation we use $y_\tau\sim0.5$ (corresponding to $t_\beta\sim50$). For $\tilde m_3$ as low as 0.5 TeV, the bound is relaxed to $|\mu|\lsim2.9\tilde m_3$. In any case, the bound from charge breaking minimum is much stronger than the bound from perturbativity, Eq.~\eqref{eq:muper}.
In the left panels of Figs.~\ref{fig:M1m3mu} and ~\ref{fig:M1m3mubranch} we show the value of $R_{\tau\mu/\tau\tau}^{\rm max}$ in the $\tilde m_3-M_1$ plane. Here we use the maximal value of $\mu$ allowed by the charged breaking constraint.

When we replace the perturbativity bound with the one from charge breaking minima, we obtain, in the bulk region,
\begin{equation}\label{eq:ccbboubul}
R_{\tau\mu/\tau\tau}\lsim10^{-4} \qquad \text{for } \left|c_{\beta-\alpha} t_\beta\right| \ll 1,
\end{equation}
to be compared with Eq.~\eqref{eq:perboubul}, and in the peninsula region,
\begin{equation}\label{eq:ccbub}
R_{\tau\mu/\tau\tau} \lsim 4\times10^{-3}\qquad \text{for } c_{\beta-\alpha} t_\beta \simeq 2,
\end{equation}
to be compared with Eq.~\eqref{eq:satub}.

The bound on $|\mu|$ becomes weaker if, instead of requiring the absence of charge breaking minima, 
we would only require that the electroweak breaking minimum is metastable with a lifetime that is 
longer than the age of the Universe. 
Following the analysis of Ref.~\cite{Hisano:2010re} we find that the bound on $|\mu|$ is relaxed by 
at most a factor of two compared to the one from Eq.~\eqref{eq:mustability}. 
(When $\mu$ saturates  the perturbative bound of Eq.~\eqref{eq:muper} the 
lifetime of the electroweak minimum is exponentially smaller than the age of the Universe.) 
Therefore, our conclusions remain unchanged.

The case of the $A$-term is somewhat different. The analysis proceeds along similar lines.
One can avoid, however, the existence of deeper minima
by taking the soft SUSY breaking parameter $M_{H_d}$ to be much larger than
the $A$-term itself (which is equivalent to the limit $c_{\beta-\alpha}\to 0$).
In this case the final result changes only by a factor of 2.

We conclude that while the MSSM can enhance $R_{\tau\mu/\tau\tau}$ by some three orders
of magnitude compared to the naive estimate of $(\alpha/4\pi)^2$, its maximal contribution
is still about two orders of magnitude below the near future experimental sensitivity.
It is interesting to note that an enhancement of the same order of magnitude can be achieved
if all possible MSSM contributions to the $h\to\tau\mu$ decay interfere constructively,
even if the trilinear scalar couplings do not saturate their upper bounds.
In either case, such an enhancement arises only in non-generic regions of the parameter space.

\section{Conclusions}\label{sec:conclusions}
The ATLAS and CMS experiments can discover the lepton-flavor violating Higgs decay $h\to\tau\mu$
if its rate is not much lower than the rate of the $h\to\tau\tau$ decay.
We examined the question of whether the minimal supersymmetric
Standard Model (MSSM) will be unambiguously excluded in case such a discovery is made.
The version of the MSSM that we analyzed has the following features:
\begin{itemize}
\item R-parity is conserved.
\item Non-renormalizable terms are negligible.
\item All couplings are perturbative. They need not obey, however, any other principle, such as flavor universality.
\item There is no charge breaking minimum that is deeper than the electroweak symmetry breaking one.
\end{itemize}
Since in this framework the $h\to\tau\mu$ decay is suppressed by an electroweak loop, while the $h\to\tau\tau$ decay proceeds at tree level, in generic points of the MSSM parameter space the LFV decay is suppressed to values orders of magnitude below the sensitivity of the LHC experiments.

When we consider only the perturbativity bounds on trilinear scalar couplings, we find very non-generic points in the MSSM parameter space that can compensate for the electroweak loop suppression. Specifically, if ${\rm BR}(h\to\tau\mu)/{\rm BR}(h\to\tau\tau)$ is discovered with a value close to the present experimental bound, the MSSM with perturbative couplings can account for it under the
following, highly non-generic, conditions: (i) The $\mu$-term is close to its perturbative bound;
(ii) There is order one lepton-flavor violation (in the $\tau-\mu$ sector)
in one of the slepton mass-squared matrices, and very small lepton-flavor violation in the other;
(iii) The bino and the sleptons have masses at the TeV scale or higher, and the higgsinos are an order of magnitude heavier;
(iv) The two lightest sleptons are quasi-degenerate;
(v) The second Higgs doublet is lighter than the sleptons and bino.

The MSSM at this corner of parameter space has, however, a charge breaking minimum that is lower than the electroweak symmetry breaking minimum. Avoiding such a minimum  (or even just requiring that the lifetime of the electroweak minimum is longer than the age of the Universe) is incompatible with the condition (i). Thus,
the $\mu$-parameter has to be smaller than the perturbative bound by about an order of
magnitude, suppressing the ratio $R_{\tau\mu/\tau\tau}\equiv{\rm BR}(h\to\tau\mu)/{\rm BR}(h\to\tau\tau)$
by at least two orders of magnitude compared to the present experimental sensitivity.

We conclude that if ATLAS and CMS establish that $R_{\tau\mu/\tau\tau}\gsim10^{-2}$,
the R-parity conserving MSSM will be excluded.

\vskip0.5cm
{\bf Note added:} While this work was in writing, Ref.~\cite{Arganda:2015uca} appeared
which also examines the possibility of a large rate for $h\to\tau\mu$ within the MSSM.
As far as a comparison is possible, we agree with their results.
In particular, the importance of the constraints from $\tau\to\mu\gamma$ is emphasized.
However, Ref.~\cite{Arganda:2015uca} calculates ${\rm BR}(h\to\tau\mu)$ and not the ratio
$R_{\tau\mu/\tau\tau}$.
Thus, the entire MSSM spectrum and mixing needs to be specified, which dictates the Higgs
total width as well as the spectrum and couplings of the $H^0$, $A^0$ and $H^\pm$ scalar
particles. With their specific choice, much smaller rates for $h\to\tau\mu$ are obtained.
In the present work, on the other hand, we are mainly interested in the possibility to
unambiguously exclude the MSSM. We thus allow for the possibility that the squark sector
strongly affects the total width and the Higgs potential, so that generic constraints
related to these aspects cannot be applied.

\subsection*{Acknowledgements}
We thank Kfir Blum, Avital Dery, Aielet Efrati, Claudia Frugiuele and Yotam Soreq for valuable discussions.
We are grateful to Roni Harnik for pointing out the potential importance of charge breaking minima to our analysis.
YN is the Amos de-Shalit chair of theoretical physics. YN is supported by the I-CORE program of the Planning and Budgeting Committee and the Israel Science Foundation (grant number 1937/12), and by a grant from the United States-Israel Binational Science Foundation (BSF), Jerusalem, Israel.

\appendix

\section{Field renormalization}\label{sec:ren}
\begin{figure}[ht]
\begin{center}
\includegraphics{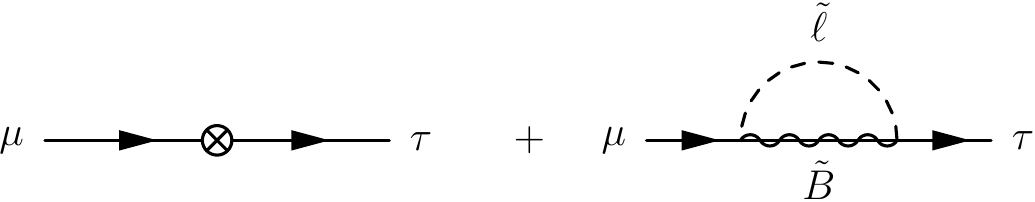}
\end{center}
\caption{Counterterm and one-loop contribution to the flavor off-diagonal
$\mu\to\tau$ two-point function.
\label{fig:TwoPoint}}
\end{figure}
To compute the transition amplitudes of physical particles, like the $h\to\tau\mu$
transition presented in Sections~\ref{sec:ae} and \ref{sec:mu}, it is convenient
to work  with a Lagrangian in which all external fields have canonically normalized
kinetic terms.
Since at the loop level this, a priori, does not hold, it is a convenient and standard
procedure to include finite parts in all the field renormalization constants of external
particles to enforce proper kinetic terms also at the loop-level.
This makes it manifest that the species of an on-shell particle cannot change in
the vacuum and is the way field renormalization constants are implemented in the
on-shell scheme.

Below we outline the procedure in the example of the $h\to\tau\mu$ decay.
In terms of Dirac fields, the relevant Lagrangian terms read:
\begin{equation}
\label{eq:barelagrangian}
{\cal L}_{\tau\tau}= \bar\tau^0i\cancel{D}\tau^0 -\left(m^0_\tau+\frac{m^0_\tau}{v^0} \frac{\sin\alpha^0}{\cos\beta^0}h^0\right)(\bar\tau_L^0\tau_R^0+ \bar\tau_R^0\tau_L^0),
\end{equation}
where the superscript ``0'' indicates bare fields and parameters.
In a chiral field theory like the MSSM the fields of different chirality are
renormalized individually.
The lepton fields are renormalized via the field renormalization constants
\begin{equation}
\ell^0_{A_i} = (Z^{A}_{ij})^{1/2} \ell_{A_j} \equiv 
\Big( \delta_{ij} + \frac{1}{2} \delta Z^A_{ij} + {\cal{O}}({\text{higher orders}})\Big) \ell_{A_j}\quad\text{with}\quad A=L,R\,.
\label{eq:fieldrenormalisation}
\end{equation}
Since we are only interested in tree-level counterterms mediating LFV transitions,
the only relevant renormalization constants at one-loop are the above
off-diagonal field renormalizations.
By inserting the expansion in Eq.~\eqref{eq:fieldrenormalisation} into
Eq.~\eqref{eq:barelagrangian}, we obtain the counterterms involving both $\tau$ and
$\mu$ fields.
We are interested in the off-diagonal terms, which are, in the limit of a massless
muon,
\begin{multline}
\label{}
{\cal L}_{\mu\tau}=
 \frac{1}{2}(\delta Z^L_{\tau\mu}+\delta Z^{L*}_{\mu\tau})  \bar\tau_Li\cancel{D}\mu_L
+\frac{1}{2}(\delta Z^R_{\tau\mu}+\delta Z^{R*}_{\mu\tau})  \bar\tau_Ri\cancel{D}\mu_R\\
-\left(m_\tau+\frac{m_\tau}{v}\frac{\sin\alpha}{\cos\beta}h\right)
\frac{1}{2}\delta Z_{\tau\mu}^R \bar\tau_L \mu_R
-\left(m_\tau+\frac{m_\tau}{v}\frac{\sin\alpha}{\cos\beta}h\right)
\frac{1}{2}\delta Z_{\tau\mu}^L \bar\tau_R \mu_L + \text{h.c.}\,.
\end{multline}
We see that the $h-\tau-\mu$ vertex is renormalised by the same two constants,
$\delta Z^L_{\tau\mu}$ and $\delta Z^R_{\tau\mu}$ as the ``mass'' vertex.
The requirement that the off-diagonal two-point vanishes at all orders in
perturbation theory fixes the two relevant constants, {\it i.e.}~the sum of the diagrams
in Fig.~\ref{fig:TwoPoint} should vanish. In our case it is sufficient
to include only the part of the two-point function proportional to $m_\tau$
because the expansion in external momenta over SUSY masses is a good
approximation:
\begin{multline}
\label{}
0\overset{!}{=}{\cal M}(\mu\to\tau)\supset-i\frac{m_\tau}{2}m_\tau\left(\delta Z_{\tau\mu}^R P_R+\delta Z_{\tau\mu}^L P_L\right)\\
+i\sum_{i=2,3}\int\frac{d^dq}{(2\pi)^d}\frac{(g_L^{\tau,i}P_R+g_R^{\tau,i}P_L)(\not q+M_1)(g_L^{\mu,i}P_L+g_R^{\mu,i}P_R)}{(q^2-M_1^2)(q^2-\tilde{m}_i^2)}\,,
\end{multline}
where $g^{\alpha,i}_{L,R}$ denote the couplings between $\ell_\alpha-\tilde\ell_i-\tilde B$.
In the case of our effective two-slepton framework, we find:
\begin{equation}
\begin{split}
\delta Z_{\tau\mu}^R&=0,\\
\delta Z_{\tau\mu}^L&=-\frac{e^{2}s_{2\theta}}{16\pi^2 c_W^2}
\frac{\tilde m_2^2-\tilde m_3^2}{m_\tau M_1}I_3(1,x_2,x_3)\,
\times
\left\{
\begin{matrix}
1                  &\text{(from $A^E$)}\\
\frac{1}{\sqrt{2}} &\text{(from $\mu Y^E$)}
\end{matrix}\right.
\end{split}
\label{}
\end{equation}
where $x_i=\tilde m_i^2/M_1^2$.

\section{\texorpdfstring{Constraints in the $\boldsymbol{\cos(\beta-\alpha)-\tan\beta}$ plane}{%
		         Constraints in the cos(beta-alpha) - tan beta plane}\label{app:ab}}

\begin{figure}[ht]
\begin{center}
\includegraphics[width=0.5\textwidth]{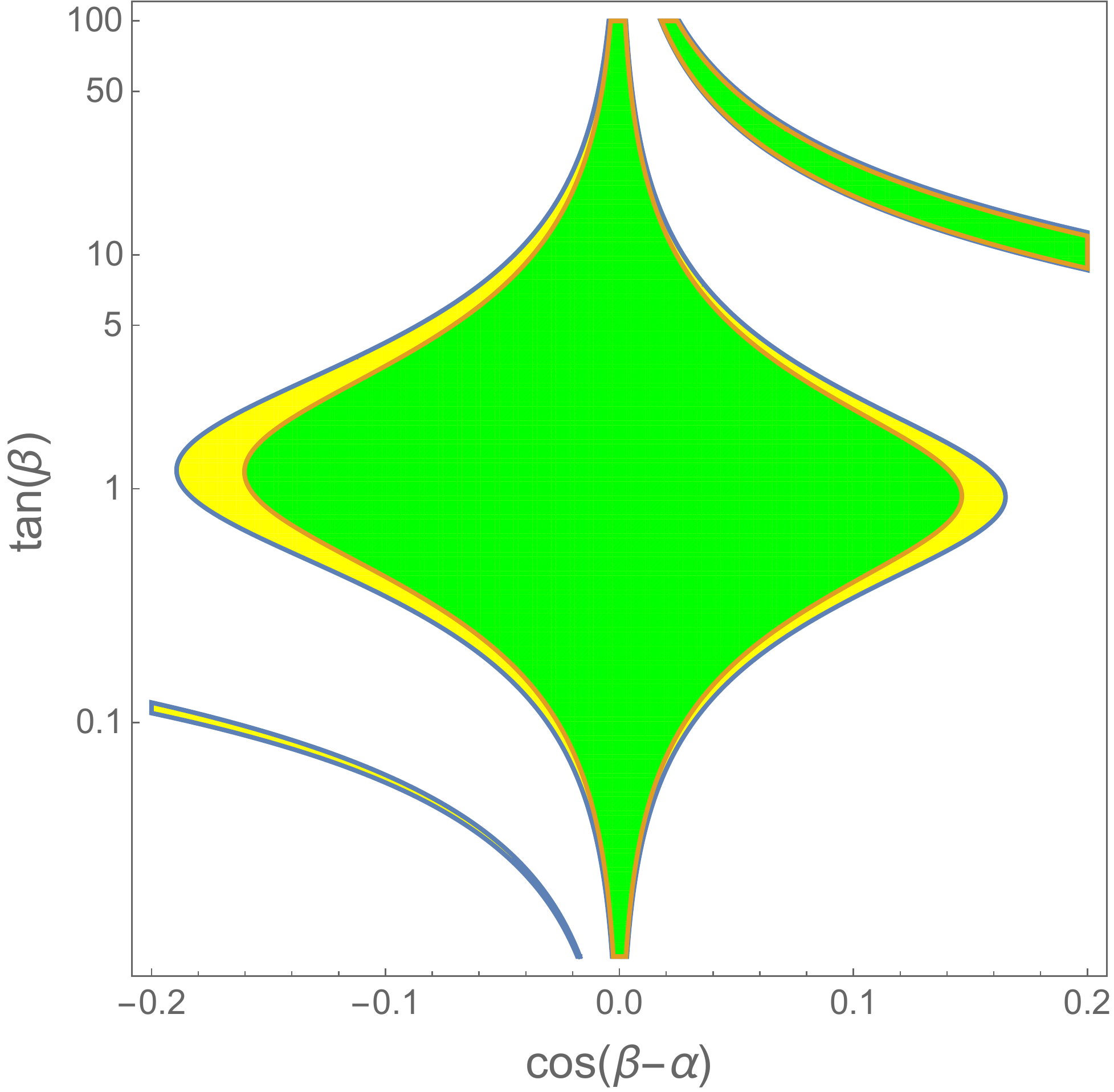}
\end{center}
\caption{Allowed region in the $\cos(\beta-\alpha)-\tan\beta$ plane.}
\label{fig:cos_ba_t_b}
\end{figure}
The couplings of the light Higgs $h$ to $VV,\gamma\gamma,\tau\tau$ and $\tau\mu$
depend on $\tan\beta$ and $\cos(\beta-\alpha)$.
We use the experimental constraints on the $h\to VV,\ \gamma\gamma$ and $\tau\tau$
modes to find the allowed region in the $\cos(\beta-\alpha)-\tan\beta$ plane.
We do not use the data on $h\to b\bar b$, as we do not commit to a specific squark sector.
We use the same set of data and procedure as employed in Ref.~\cite{Craig:2015jba}.
To do so, we fix $\tan\beta$ and $\cos(\beta-\alpha)$ to all orders in perturbation theory
through:
\begin{equation}\label{eq:defab}
\begin{split}
g_{hVV}/g_{hVV}^{\rm SM}&\equiv\sin(\beta-\alpha),\\
g_{h\tau\tau}/g_{h\tau\tau}^{\rm SM}&\equiv\sin(\beta-\alpha)-\cos(\beta-\alpha)\tan\beta.
\end{split}
\end{equation}
Note that this definition of $\tan\beta$ differs at the loop level from the standard
definition of $v_u/v_d$.
With this definition, our results should reproduce, to a good approximation,
the allowed region for 2HDM Type II obtained in Ref.~\cite{Craig:2015jba},
and indeed they do.

The allowed region in the $\cos(\beta-\alpha)-\tan\beta$ is presented in Fig.~\ref{fig:cos_ba_t_b}.
We refer to the central allowed region, which includes the point $(0,1)$ as ``the bulk region''.
We call the branch in the upper right corner ``the peninsula region'', a term used in Ref.~\cite{Craig:2015jba}.
It corresponds to $\sin(\beta-\alpha)$ not far from $1$, and $\cos(\beta-\alpha)\tan\beta$ not far from $2$.
In this way, the $hVV$ and $h\gamma\gamma$ couplings are close to their SM values, while the
$h\tau\tau$ coupling has the same absolute value but opposite sign.

\section{\texorpdfstring{$\tau\to\mu\gamma$}{
         tau->mu gamma}\label{app:tmg}}
To obtain the rate for $\tau\to\mu\gamma$, we integrate out all the heavy degrees of freedom,
and match to the effective Lagrangian
\begin{equation}
{\cal L}_{\rm eff}=-\frac{e}{2}C_\gamma \overline{\mu}_L\sigma^{\mu\nu}\tau_R F_{\mu\nu}+\text{h.c.}\,.
\end{equation}
The Wilson coefficient $C_\gamma$ gives the rate:
\begin{equation}
\frac{{\rm BR}(\tau\to\mu\gamma)}{{\rm BR}(\tau\to\mu\nu\bar\nu)}
=\frac{48\pi^3\alpha}{G_F^2m_\tau^2}|C_\gamma|^2.
\end{equation}
The relevant experimental results are \cite{Hayasaka:2007vc,Agashe:2014kda}
\begin{equation}
\begin{split}
{\rm BR}(\tau\to\mu\gamma)&<4.5\times10^{-8}\ {\rm at}\ 90\%\ {\rm C.L.},\\
{\rm BR}(\tau\to\mu\nu\bar\nu)&=(17.41\pm0.04)\times10^{-2}.
\end{split}
\label{}
\end{equation}
With this experimental input the bound on the Wilson coefficient reads:
\begin{equation}\label{eq:cgammabound}
|C_\gamma|<3.1\times10^{-9}\ {\rm GeV}^{-1}.
\end{equation}

In the cases of interest to us, we have an effective two-slepton framework,
with the mass eigenstates $(\tilde\ell_2,\tilde\ell_3)$.
We denote the mixing matrix for bino-slepton-lepton couplings by $\tilde U$.
We obtain:
\begin{equation}
C_\gamma=-\frac{\alpha}{8\pi c_W^2 M_1}\sum_{i=2,3}\tilde U_{i\mu_L}\tilde U_{i\tau_R}^*
\frac{1-x_i^2+2x_i\log x_i}{(1-x_i)^3},
\end{equation}
where $x_i\equiv \tilde m_i^2/M_1^2$. Then, the upper bound on $|C_\gamma|$ (\ref{eq:cgammabound}) translates into
\begin{equation}
\frac{130\ {\rm TeV}}{M_1}\left|\sum_{i=2,3}\tilde U_{i\mu_L}\tilde U_{i\tau_R}^*
\frac{1-x_i^2+2x_i\log x_i}{(1-x_i)^3}\right|<1.
\end{equation}

For the case of LFV from the $A$-terms discussed in Section \ref{sec:ae}, we have $\tilde U_{2\mu_L}\tilde U_{2\tau_R}^*=-\tilde U_{3\mu_L}\tilde U_{3\tau_R}^*=1/2$. For the case of LFV from the $\tilde m^2_L$-terms discussed in Section \ref{sec:mu}, we have $\tilde U_{2\mu_L}\tilde U_{2\tau_R}^*=-\tilde U_{3\mu_L}\tilde U_{3\tau_R}^*=1/(2\sqrt2)$. Thus the contribution to $|C_\gamma|^2$ is smaller by a factor of $1/2$ in the latter case, compared to the former, and the lower bounds on the spartner masses from $\tau\to\mu\gamma$ are correspondingly weaker. In either case, taking into account that the large slepton mixing entails quasi-degeneracy between the two sleptons, we expect the lower bound on the bino and/or slepton masses to be of order 10 TeV.
The numerical impact is shown in the relevant sections.

\section{\texorpdfstring{$h\to\gamma\gamma$}{
         h->gamma gamma}\label{app:hgg}}
Within the MSSM, the $h\to\gamma\gamma$ decay rate is given by
\begin{equation}\label{eq:hgg}
\Gamma(h\to\gamma\gamma)=\frac{G_F^2\alpha^2 m_h^3}{128\sqrt2\pi^3}\left|\sum_f c_f A_{1/2}(\tau_f)
+c_w A_1(\tau_W)+A_{\tilde\ell}\right|^2,
\end{equation}
where $\tau_{f,W}\equiv m_h^2/4m_{f,W}^2$,
\begin{equation}
\begin{split}
A_{1/2}(\tau)&=2[\tau+(\tau-1)f(\tau)]/\tau^2,\\
A_1(\tau)&=-[2\tau^2+3\tau+3(2\tau-1)f(\tau)]/\tau^2,
\end{split}
\end{equation}
and
\begin{equation}
f(\tau)=\left\{\begin{matrix}\arcsin^2\sqrt{\tau}&\tau\leq1,\\
-\frac14\left[\log\frac{1+\sqrt{1-\tau^{-1}}}{1-\sqrt{1-\tau^{-1}}}-i\pi\right]^2 & \tau>1.
\end{matrix}\right.
\end{equation}

The first term in Eq.~\eqref{eq:hgg} comes from the SM fermion loops and the second from the $W$-boson loop, but with the MSSM couplings:
\begin{equation}
\begin{split}
c_t&=\frac43(s_{\beta-\alpha}+c_{\beta-\alpha}/t_\beta),\\
c_b&=\frac13(s_{\beta-\alpha}-c_{\beta-\alpha}t_\beta),\\
c_\tau&=s_{\beta-\alpha}-c_{\beta-\alpha}t_\beta,\\
c_w&=s_{\beta-\alpha}.
\end{split}
\end{equation}
The third term comes from the slepton loop and, for $\tilde m_{2,3}\gg v$, is given by
\begin{equation}
A_{\tilde\ell}=\frac{1}{6\sqrt2}\left(\frac{1}{\tilde m_2^2}-\frac{1}{\tilde m_3^3}\right)\times\left\{
\begin{matrix}-s_\alpha v A_{\mu\tau} &\ ({\rm from}\ A^E),\\
(c_{\beta-\alpha}+s_{\beta-\alpha}t_\beta)m_\tau\mu &\ \ ({\rm from}\  \mu Y^E).\end{matrix}\right.
\end{equation}
%

\bibliographystyle{utphys}
\bibliography{references}
\end{document}